# A Novel Adaptive Caching Mechanism for Video on Demand System Over Wireless Mobile Network

Saleh Ali Alomari

Faculty of Sciences and Information Technology, Jadara University, Irbid, Jordan

*Abstract*

*Video on Demand (VOD) system over the wireless mobile network is a system that provides video services to mobile clients. The main problem with these systems is the high service delay where the mobile clients have to wait to view their favorite movie. The importance of this paper is based on finding a solution on how to reduce the delay time in the VOD system. This paper introduces a novel caching mechanism named Proxy Server Cache mechanism to tackle the issue of service delay. This delay happens when the broadcasting phase that is related to the first segment is missed by a client from the current broadcasting channels. In this mechanism, the video's first segment is stored on a server of a stationary proxy type. The delayed clients will directly acquire the first segment from the proxy server instead of waiting for the following broadcasting channel pertaining to the first segment. The proposed scheme ensuresobtaining the first segment from mobile clients when they arrive. Additionally, the performance of the proposed scheme is validated by applying the VOD system, which can involve the balancing mechanism to retain particular requests through to the local proxy server to provide a fair dissemination for these requests. The obtained result confirms that the proposed scheme reduces the time delay of the system in comparison with the best existing schemes. The results of the average time delay in the Proxy-Cache scheme is 179.2505 milliseconds when 10 clients arrive each minute (Client/minute), the average time delay is 140 milliseconds when the video lengths are 30, 60 and 90. Meanwhile, the failure probability for obtaining the first segment of the video remains zero when the number of arrived requests is set to2, 4, 6, 8 and 10.*

## 1. Introduction

The Video on Demand (VOD) system is considered to bean emerging system, which allows a user to be able to view any video anywhere and anytime. This system can be implemented under several approaches according to the technique that is used to design the VOD system. The design of this system is classified into three approaches, which comprises the Client/Server, Peer-to-Peer (P2P) and Periodic Broadcast (PB) [1-3].Some limitations are included within these approaches. In the first approach, the Client/Server approach is incompatible with the Mobile Ad Hoc Network (MANETs), the reason behind that the limitation of the wireless bandwidth. Further, this limitation produces many issues when an increment occurs on the number of requests. The second approach implements the P2P approach such that it is not suggested to transmit long videos over more than a single wireless hop due to the inefficient energy and bandwidth being used through the requested process. In the third approach, the Periodic Broadcast approach is efficiently used in order to tackle the bottleneck problem through the server and client, and meantime, through the vulnerable service of the P2P approach. Compared to the previous approaches, the periodic broadcast approach is considered an efficient approach based on its bandwidth since any number of mobile clients can be served in efficient manners [4-6]. In particular, the client can switch to one or multiple server channels and could provide the ability for the client to view the video seamlessly. Accordingly, the number of requests will not put an impact on the server bandwidth. The entire broadcasting protocols such as those protocols mentioned in [7-14] classify the video into different numbers of segments and frequently broadcast these segments throughout the number of channels. Mobile wireless technologies allow users to enjoy viewing what they wantanytime and anywhere within the coverage area. After developing and utilizing the VOD system as an interactive multimedia system, a lot of practical applications such as movies-on-demand, video conferencing, distance learning, interactive video games, etc., can be implemented due to the advancement of the wireless technology. Some of these applications are utilized to make users enjoy and provide entertainment services, such as playing an online game or viewing an online video of their interest wherever they are. Universities can also apply such a system throughout their campuses in order to enable students to view the videoin advance when the video is being recorded from seminars/lectures who are unable to attend them.

Current trends have drastically made an impact on VOD services due to the deployment of various types of network infrastructures and the availability of different types of mobile devices. These services have become interactive and efficient multimedia services, which assists clients to playback any large collections of videos anytime within public communication networks in a simple way after waiting for a short time. Hence, these clients could effectively use many interactive services and could be able to seamlessly download as many videos as possible at the same time [15]. These video files are stored in a set of servers related to the central video and are disseminated over networks that are of high-speed communications in order to geographically make clients be dispersed. Mobile computing devices and wireless technologies provide more convenient and flexible services to clients to enjoy viewing more efficient videos. Additionally, such technologies provide flexible services derived from a distant video server when they are seamlessly moving within their wireless network transmission coverage. The proposal of appropriate system architecture relies on the server's location, available storage, communication systems and other related factors. The main elements of the VOD system comprise user display equipment, networks and servers. The server stores a large number of videos and broadcasts them all for users. Users request the videos and view them by applying the display equipment into the network, such as PDA's, iPad, smart phones and many other devices. The features of the VOD system that involve the requirements of the high bandwidth, QoS-sensitive service and long-lived session make the design of the VOD system be more challenging. The reason behind this refers back to the nature of the video data. The data of a video involves the real-time data that should meet the requirements of a particular Quality of Service (QoS). Additionally, the issues emerge when fast videos can be viewed once a client requests a desired video. Consequently, this case refers to the time delay, which is called the average period of time for which a client should wait until being served. It selects an end-to-end time, which is considered to represent the difference between the times of requests when a packet transfers from the source (server) to the time of arrival to the user's destination. There exist several elements that influence the problem of the time delay. The choice of the system architecture is considered to be one of the elements, which puts an impact on the overall performance, particularly, the storage availability, the servers' locations, the protocol of communication systems and other related factors that are likely capable to contribute to the time delay.

The following sections are structured as follows. Section 2 presents the related research. In Section 3, the cache proxy server mechanism is shown. In Section 4, the broadcasting techniques along with its characteristics for the proposed scheme are provided. The proposed novel caching mechanism scheme is thoroughly discussed in Section 5. The scenario of the playback pertaining to the proposed scheme for the VOD system is introduced in Section 6. The experimental environments and system parameters are presented in Sections 7 and 8. The experimental results and discussions are given in Section 9. Finally, the conclusions are all drawn in Section 10.

## 2. RELATED RESEARCH

The Video on Demand (VOD) system is improved by using multicast or broadcast schemes where most of the multicasting protocols [16-18]are reactive in the sense that they transmit data in response to users' requests. Multicasting protocols attempt at making users share the same stream of data as much as possible. While some of the multicast approaches can provide an immediate service and save server bandwidth by avoiding unnecessary transmission of data, they are subject to data loss and cannot guarantee an on-time delivery of data if users' requests are extremely high. Broadcasting protocols can address this problem by periodically transmitting the video segment in a proactive way and by ensuring the service latency within a certain amount of time. In order to provide different VOD services, many VOD Periodic Broadcasting (PB) protocols are proposed, such as the Staggered Broadcast [19], skyscraper [20] the Harmonic Broadcasting [21]and Fast Data Broadcasting [22], etc. The main idea for the PB protocols is to partition the video into several segments and broadcast each segment periodically on dedicated server channels. While the user is playing the current video segment, it is guaranteed that the next segment is downloaded on time and the whole video can be played out continuously. In this case, the user will have to wait for the occurrence of the first segment before they can start playing the video. The user waiting time usually represents the first segment's length. The broadcasting method of the server is considered another element, which contributes to the waiting time, including the number of concurrent users that are supported by the VOD system. Broadcasting is based on the scenario of handling clients who are viewing many different portions related to the same video at any provided time. The server must have an effective broadcasting mechanism in order to broadcast the video to as many users as possible in a form of a simultaneous fashion with according to the requirements of a stringent delay.

In general, the broadcasting technique is considered to be a method that disseminates a video from a server to several simultaneous users. Broadcasting schemes are proposed to efficiently be implemented with many different network infrastructures [23] including Local Area Network (LAN), direct broadcast satellite and cable TV. The broadcasting schemes aim at providing efficient ways of reducing the delay time among clients. As previously indicated this time is based on the interval time between the requests for a video and receiving that video from the start. Over the past two decades, many researches use the VOD system for providing optimal services to mobile ad hoc devices. The MobiVoD system [4] [5] is considered to be one of the up-to-date VOD system, which provides video services to different homogeneous devices within the homogeneous network. This system allows mobile clients to view their desired movies. Additionally, the system is comprised of three major components, which are Mobile Clients, Local Forwarder (LF) and Video Server. Basically, all video files are stored on the main server. The LF indicates to the number of disseminated stationary devices that is used as a relay for the video to mobile clients through broadcasting video segments to its coverage service area. The server does not transmit the video by using the wireless technique to different wide coverage areas. Sub sequently, the LF is used to enlarge the transmission around the coverage area. The Staggered Broadcast (SB) protocol is repeatedly broadcast video segments through to mobile clients on differentnumbers of wireless broadcasting channels. The SB is known to be the most effective selection for broadcasting the video through to the mobile devices due to the limitation of the mobile devices [4]. Several researchers propose many different caching techniques that could solve the problem of time delay. In [24] introduced a novel caching technique caching technique that can be applied to numerous applications of the client-server interaction model for an optimized storage and transmission of data, which can support mainly three data types, images, video and XML document. Cached elements are accessible via a local proxy by neighboring nodes requesting similar data, the proposed caching discussed for the scalability and increased lifetime of mobile ad hoc networks. In [4][15], several types of caching, such asRandom Caching, DSC Cashing, PoR Cashing and All Caching, are proposed. Random Caching and All Cache schemes are also proposed in [4-6]. In the All-Cache scheme, the entire clients of the LF coverage must store the first segment. However, a random number of clients have to store the first segment in the Random Caching. In both schemes, the late clients can cache the first missed segment from its neighbor who already stores the first segment into their buffer. The advantage of this caching scheme is the saving process on the caching space and the disadvantage is that it contains a higher service delay. The DSC-Cache scheme is proposed in [4] where the current Client X can obtain the first segment from Client Y within the same transmission area, and Y must not transmit the segment to any other clients at the same time. Nonetheless, if Client Y does not appear within the transmission area of Client X, after that, Client Y finds Client Z who shares the same transmission area with Client X and the same client (i.e. Client Z).Client Y transmits the first segment to Client X through Client Z. In [17], the PCSB protocol is produced in order to overcome the encountered issue of the late clients who loses the first segment pertaining to the current broadcasting channel. In order to solve this issue and provide the possibility for the client to obtain the video segment (the first segment) without the need to wait for the following broadcasting channel, the PCSB assists the clients to directly obtain it from the Pool of RAM (PoR) of an available Media Forwarders (MF), where it utilized as a process for storing the first segment of all videos that exist in the PoR of the MF once these videos are broadcasted to the viewers. The PCSB scheme can reduce the time delay by ensuring that the late client can obtain the first segment once it arrives. In [25], the cache is enhanced by providing the on-going video streaming through to the Relay Station (RS) in order to reduce the delay time. The Proxy Prefix Caching for Multimedia Streams is introduced in[26]in order to minimize the network resource requirements and the user-perceived latency through the Internet where a proxy is utilized to store initial frames of video or audio. The proxy directly transmits the data to the client whenever it is requested. Additionally, in [27], the peer-to-peer caching techniques use the Radius Based Binary Search Algorithm (RBBSA) and mobile management in order to solve the problem of mobility constantly, which makes an interruption of the video packets transmission." The results show that the hit ratio of the P2P caching techniques can increase the caching and reduce the cache distance where the cluster head for the node mechanism can provide the video to another node such that the node acts as a server. In[28], initial caching schemes (i.e. PSCM scheme) are proposed to overcome the service delay time issue through mobile clients where this issue is caused based on the use of a broadcasting technique over the mobile VOD system of the Ad Hoc network. This issue is accursed when a client loses the broadcasting phase of the first segment that is transmitting from the broadcasting channels. In the first segment of the PSCM scheme, the entire videos are stored in a stationary server. The late clients immediately request the first segment from the proxy server instead of waiting for the following broadcast of this segment. Moreover, there exist several proposed caching techniques such as those indicated in[29-35].

## 3. SYSTEM OVERVIEW FOR THE CACHE PROXY SERVER MECHANISM

The new caching mechanism is proposed to reduce the start-up delay through VOD systems where the first segments of the whole broadcasted videos are efficiently stored in the local Proxy Server. If the client arrives when the LF broadcasts its first segment, the clients will not be able to obtain it. Hence, the client shall wait untilthe following broadcasting channel could broadcast the first segment. In order to tackle this problem, installing a stationary proxy is suggested in this paper, and which is also suggested to be called as the Local Proxy Server. This type of server ensuresthat late mobile clients could certainly obtain the first segment once they arrive with a reduced start-up time delay in comparison with other related techniques. In this paper, the VOD system is categorized into four components, which include the number of mobile clients, LFs, the main server and the Local Proxy Server (LPS). The main server is a type of server that is being utilized to store several types of video files as shown in Figure 1. The LF is considered a stationary device that is being utilized to relay the videos through to their wireless coverage area. The LPS is considered a stationary server that is being utilized to store the first segment of each

International Journal of Computer Networks & Communications (IJCNC) Vol.10, No.6, November 2018

broadcasting video for the purpose of reducing the delay. Clients are considered to be mobile devices that can view and receive videos.

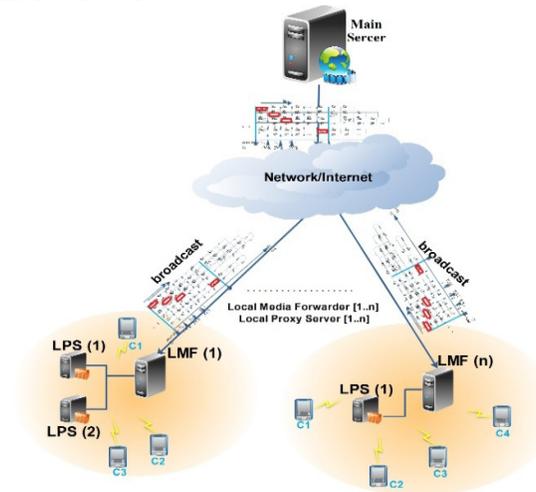

Figure 1. The overall system of the newly proposed caching scheme

## 3.1 FEATURES OF THE LOCAL FORWARDER (LF)

In the proxy server cache scheme, the main task of the LF is to manage the VOD system. This task includes utilizing the Staggered Broadcasting (SB) protocol for broadcasting the videos over multiple numbers of logical channels, assigning the clients to a proper broadcasting channels, passing the first segment of the videos being broadcasted to the LPS and managing the requests of the late clients over the LPSs to balance the load of the LPSs. Figure 2, shows the scenario when the LF broadcasts a video over its logical channels. Each client requests the video directly from the LF by tuning to one of these channels.

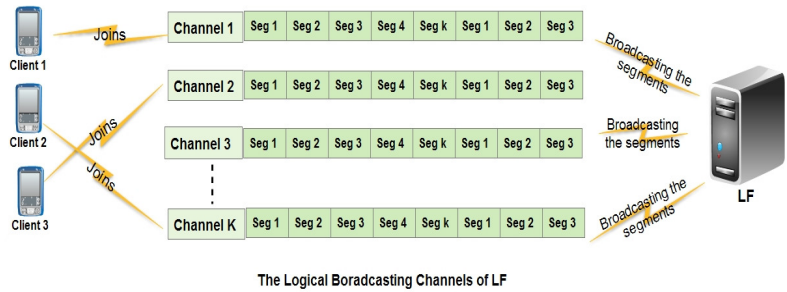

Figure 2. Logical broadcasting channels of the LF

Figure2 illustrates the logical broadcasting channels of the LF where clients receive the video form the LF when they arrive at a suitable time of broadcasting the first segment. Each client joins one particular channel. This channel is assigned to the client depending on the client arrival time. Moreover, this figure depicts that Client 1 joins Channel 1of the LF, Client 2 joins Channel K and Client 3 joins Channel 2. Each client starts streaming the video after joining their channels.

## 3.2 FEATURES OF THE LOCAL PROXY SERVER (LPS)

In the proposed scheme, a proxy server is a device that acts as a caching buffer to speed up the access through to the first segment of the videos. This device is recognized as one of the most effective schemes for alleviating server bottlenecks, reducing the network traffic and minimizing

International Journal of Computer Networks & Communications (IJCNC) Vol.10, No.6, November 2018

the access time. This technique is mostly being used to reduce the network traffic between the client/server networks where the proxy server is always located between the client and the server in order to hold the data which is frequently being used. In the proposed scheme, the LPS is a stationary device that is used for streaming the first segments of the videos to multiple clients at one time. Adding LPS to the system ensures obtaining the first segment immediately since the first segment is available anytime. Consequently, this saves time for clients and makes the service of video broadcasting as an actual on-demand service. In the proxy server caching scheme, the LPS allows clients to obtain the first segment when they arrive at time ($T_0$+s). This time represents the arrival time of the client after broadcasting the first segment when the LF starts. The LPS is exploited for handling the timing issue in a way that allows late clients to obtain the first segment even after missing the current first segment that is broadcasted by the LF.

### 3.3 FEATURES OF THE MOBILE CLIENT DEVICES

Client devices include portable computers and phones that can make use of the video service. These devices are required to be capable of receiving and viewing videos so that a seamless video streaming can be provided to the users. In this paper, the VOD system is designed to provide the VOD service for two types of mobile phone devices that are 3G and 4G mobile phones. These types of devices are capable of rendering the video (receiving the packet, decoding the packet and displaying the video). These devices can connect through the Wi-Fi (IEEE 802.11) technique. The memory of the mobile phone is logically divided into two buffers. The first one is the initial buffer and the second is the per-fetch buffer. The initial buffer is used as a temporary buffer while streaming the first segment of the video from either the LF or the LPS, whereas the per-fetch buffer is used as a temporary buffer when streaming the rest of segments from the LF. The size of the initial buffer is equal to the first segment of the video, whereas the size of the per-fetch buffer is equal to the missing portion of the broadcasted data. Figure 3 shows a mobile client's phone.

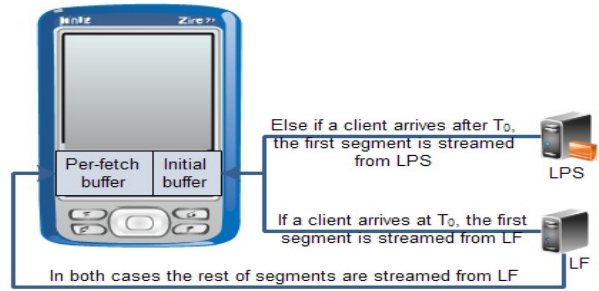

Figure 3. The buffer at the mobile client device

The interconnection between the LF and the main server can either serve an infrastructure-based wireless network or a wired LAN. The devices LF, LPS and Clients are entirely equipped with a Wireless Network Interface Card (WNIC) to provide communications with each other within the IEEE 802.11b technique. The video file is classified into a number of equal-sized segments, which are broadcasted to LFs via the main server. The LF distributes the video within its coverage area based on broadcasting such segments within a number of transmitting channels. One of these logical channels are communicated by the clients who view the video as illustrated in Figure1, which shows the LF's logical broadcasting channels, and which demonstrated how the clients communicate with these channels. Let define video k with Qth quality, it is denoted as (VkQ), which is encoded at a rate $S_{kQ}^{rate}$ thatis denoted as follows: $S_{k1}^{rate}$, $S_{k2}^{rate}$, $S_{k3}^{rate}$, …, $S_{kQLk}^{rate}$. We first consider how to determine whether the video is stored in the LMF or not. It is assumed that $p_{rob}R_j$ denotes the probability of the users who are requesting the **VkQ∀k** where

, $\sum_{Q=1}^{QLk} p_{rob}R_j = 1$. In the proposed system, the LMF simply stores the most popular videos in order to maximize the caching hits. We define the media forwarder map as $MF_{KQ}$, which is used to describe the subsets of video replicas within its cache. The $MF_{KQ}$ is set to 1 if the VkQ is in the media forwarder. Otherwise, it is set to 0. Therefore, the cache hits the optimization problem, which can be expressed as follows Equation 1 and Equation 2:

$$\sum_{k=1}^{k}\sum_{Q=1}^{QL_k} P_{rob}K * P_{rob}R_j * MF_{KQ} \qquad (1)$$

Where, $P_{rob}K$ denotes the probability of the video and $P_{rob}R_j$ denotes the probability of the user's request.

$$\sum_{k=1}^{k}\sum_{Q=1}^{QL_K} S_{KQ} * MF_{KQ} \leq S_{MF} \qquad (2)$$

Where, $S_{KQ}$ denotes the size of the video k encoded in Qth quality (bits) and $S_{MF}$ denotes the size of the media forwarder.

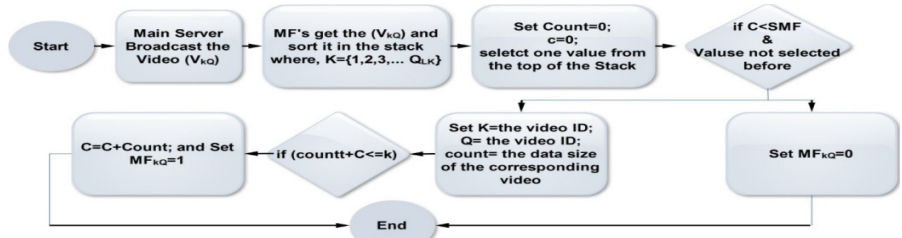

Figure 4. The procedure of determining which video should be cached in the LMF

Based on Figure 4, VkQ is broadcasted from the main server to the LMFs. After that, the VkQ is sorted by the LMFs in an ascending order into the stack based on the popularity $P_{rob}K * P_{rob}R_j$. In the stack, each value contains a ($VKQ, P_{rob}K * P_{rob}R_j$) pair. From the top of the stack, one value is selected by each loop. If the cache is high inside the LMF to accommodate the related video content, the VkQ for this video is set to value "1". Otherwise, the process continues and the following value is selected by this process from the stack until the entire caching space is allocated. Once VkQ is found by maximizing the efficiency of the cache, the fraction of requests can be identified. This fraction rises to the main server for the dedicated streams. Since the LMF is already saturated by a number of requests, the remains of the request get through the LPS where Equation 3 can compute the arrival rate of these requests. Since the deliveries of the multiple qualities of the video streams are at different data rates from the main server, LMF and LPS of the mobile clients, Equation 4 can calculate the average streaming rate.

$$\lambda_{stram}^{rate} = \lambda \left(1 - \sum_{K=1}^{K}\sum_{Q=1}^{QL_K} P_{rob}K * P_{rob}R_j * V_{KQ}\right) \qquad (3)$$

Where $\lambda_{stram}^{rate}$ denotes the arrival rate of the dedicated stream (request/second) of the broadcasting.

International Journal of Computer Networks & Communications (IJCNC) Vol.10, No.6, November 2018

$$AS^{rate} = \frac{\lambda}{\lambda_{stram}^{rate}} \sum_{K=1}^{K} \sum_{Q=1}^{QL_K} P_{rob}K * P_{rob}R_j * S_{KQ}^{rate} * \overline{MF}_{KQ} \qquad (4)$$

Where $AS^{rate}$ denotes the average stream rate of the detected stream (bit/ second) of the broadcasting, $S_{KQ}^{rate}$ denotes the streaming rate of Video k having a Qth quality level (bits/ second) and $\overline{MF}_{KQ}$ denotes the complement of $MF_{KQ}$. The scalability issues can be raised by the main server. The reason for this is that it is considered to be the bottleneck of the system itself when numerous numbers of video streams are being served. Therefore, we particularly concentrate on the performance of both the LMF/LPS and the main server to provide VOD services.

The available bandwidth between the main server and LMF is referred to as b. The number of video streams $N_{vi}$ is supported by both of them at the same time according to Equation 5. Moreover, we assume that the service time (**T**) of each video stream is exponentially distributed with the service rate $\mu = \frac{1}{T}$ by considering the varying length of different videos.

$$N_{vi} = \frac{b}{AS^{rate}} \; where, v_i = \{1,2,3,\dots,K\} \qquad (5)$$

As shown in Equation 6, the blocking probability is formalized. If the bandwidth from the media forwarder to the clients is large enough and no requests are blocked, the overall of the blocking probability of the system is given by Equation 7.

$$P_{rob}b^r = \frac{\left(\frac{\left(\frac{\lambda_{stram}^{rate}}{\frac{1}{T}}\right)N_{vi}}{N_{vi}}\right)}{\sum_{z=0}^{N_{vi}}\left(\frac{\left(\frac{\lambda_{stram}^{rate}}{\frac{1}{T}}\right)z}{z}\right)} \qquad (6)$$

$$P_{rob}OA^{rate} = \frac{\lambda_{stram}^{rate} * P_{rob}b^r}{\lambda} \qquad (7)$$

Where, $P_{rob}OA^{rate}$ denotes the overall blocking probability of the system and λ denotes the system arrival rate (request/ second).

## 4. BROADCASTING TECHNIQUE FOR PROXY SERVER CACHE MECHANISM

The main server transmits the videos through to the LF, which uses the SB protocol as a broadcasting technique that could broadcast the video through to its clients such that this video is divided into a number of equal-sized segments. These segments are regularly broadcasted on a number of logical channels based on the scheduled time of the SB protocol where the number of logical broadcasting channels is equivalent to the number of segments. The number of logical broadcasting channels is selected based on the bandwidth pertaining to the transmission media. This bandwidth is partitioned into K logical broadcasting channels. Every video is broadcasted through K logical channels for the purpose of increasing the opportunity for the clients to efficiently obtain the video at a determined time. The bandwidth that is needed by the system is constant and the system can meet any number of requests. This explains the use of the broadcasting technique within the system. In the Proxy Server Cache mechanism, the entire video is partitioned into equal K size segments$(Seg^1, Seg^2, Seg^3, ..., Seg^K)$. The duration of every segment is Di = V/K. The number of each logical broadcasting (Channel$_i$) must be ranged as(Channel$_i$ = $1 \leq i \leq K$).I suggested that the provider bandwidth is set toPb*K for the second video and beyond that. This bandwidth is frequently divided into physical channels (Channel$_i$) by broadcasting the video beginning with (Seg$^1$) and ending at the video (Seg$^K$) within a transmission rate (Tr) that is equal to the rate of the playback (Pb) (see Figure4). Client_x can join Channel1 and wait for the start of the first segment (Seg$^1$) in order to download it and play it back. Accordingly, Client_xtransfers to the following segment (Seg$^2$) for the playback. This process is frequently being performed for the subsequent segments until the last segment (Seg$^K$) is downloaded from Channel1. Equations 1 and 2 follow the definition of Equations 8 and 9.

$$Di = \frac{V}{K} \qquad (8)$$

$$V = \sum_{i=1}^{K} Di \qquad (9)$$

Where,$Di$ denotes the duration of each segment, V denotes the length of the video and K denotes the number of the channels. The equation that is used to determine the number of logical channels (k) is given by Equation 10.

$$T_r * K * N_{vi} \leq b \textbf{where}, i = \{1,2,3,...,n\} \qquad (10)$$

Where,Trdenotes the bandwidth transmission media, N*vi*denotes the number of videos, K denotes the number of logical channels and b denotes the bandwidth.

Figure 5 illustrates the specifics of the SB mechanism for the video broadcasting within the logical channels where Channel 1 begins broadcasting the video segments in a sequential way. After some time (equals to the time of the segment), Channel 2 also begins broadcasting the segments in the same way as occurring in Channel 1. The same mechanism is performed for the other remainingchannels. The video can be directly viewed in the SB technique once the client arrives at time T0 when the video is broadcasted on the LF channel. T0 represents the time in which the start of the video broadcasting of the first segment takes place. The client should wait for a period of time until the following broadcasting of the first segment begins on another channel, if this client arrives at T0+s of the first segment broadcasting.

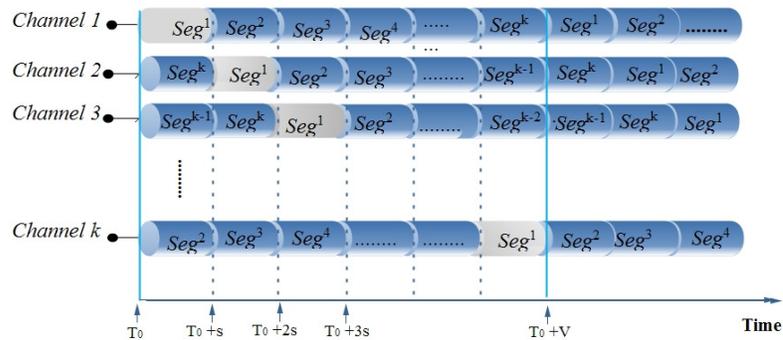

Figure 5.The broadcasting mechanism of the proxy server caching scheme

International Journal of Computer Networks & Communications (IJCNC) Vol.10, No.6, November 2018

By employing the caching and broadcasting techniques, the enhancement in the performance can be gained. Apart from storing the popular videos in the LMF, some popular videos are also broadcasted to the clients over the backbone network and over the LPS. For example, a low-quality video can be delivered through the broadcasting channels, while the higher encoded version of the same video is transmitted to the clients through the dedicated streams. It is necessary to determine which video can be delivered over the broadcasted channels. Since our goal is to improve the overall performance of the system, the broadcasting and caching techniques. In general, any efficient protocols, such as in[6][15][36][37],can be applied to the framework as the broadcasting bandwidth and caching requirement are based on the transmission schedule and user bandwidth constraints. The LPS channels are denoted as KLPS (KLPS= {$C_{hannel}^1$, $C_{hannel}^2$,…$C_{hannel}^k$}) as the number of channels are required for the LPS to broadcast a video such that the start-up delay is insensitive to the clients. It is also assumed that each receiver is equipped with enough buffers to implement the efficient broadcasting protocol. In order to determine which popular video should be sent over the broadcasting channels, we use $X_{KQ}$ to check as to whether the $V_{KQ}$ is already broadcasted or not. The consumption bandwidth for broadcasting is calculated according to Equation 11 as follows:

$$b^{brod} = \sum_{K=1}^{K} \sum_{Q=1}^{QL_K} S_{KQ}^{rate} * K^{LPS} * \overline{MF}_{KQ} * X_{KQ} \qquad (11)$$

Where, $b^{brod}$ denotes the bandwidth that is required for broadcasting, $K^{LPS}$ denotes the number of the channels, $\overline{MF}_{KQ}$ denotes the complement of the streaming rate if require video k have Qth quality level per-bits ($MF_{KQ}$).

Similarly, $X_{KQ}$ is selected for the broadcasting channels in the LMF caching according to their popularity. For example, based on the previous explanation, the videos are sorted depending on their popularity where the most popular video, which is the "first video in the stack" is broadcasted. Assume that the broadcasting bandwidth is preserved; it is found that the video $X_{KQ}$ with the broadcasting bandwidth does not exceed the capacity of the existing bandwidth. This implies that the required broadcasting bandwidth needs to be less or equal to the reserved broadcasting bandwidth (b<= preserved). This occurs due to some replicated videos, which are being broadcasted. Equation 12 demonstrates the arrival rate for the dedicated channels. This rate is equal to the arrival rate to the system minus the arrival rate to the LMF and the arrival rate to the broadcasting channels itself. The average streaming rate of the dedicated channels can thus be found by Equation 13.

$$\lambda_{stram}^{brod} = \lambda \left( 1 - \sum_{K=1}^{K} \sum_{Q=1}^{QL_K} P_{rob}K * P_{rob}R_j * MF_{KQ} - \sum_{K=1}^{K} \sum_{Q=1}^{QL_K} P_{rob}K * P_{rob}R_j * \overline{MF}_{KQ} * V_{KQ} \right) \qquad (12)$$

$$AS^{broad} = \frac{\lambda}{\lambda_{stram}^{broad}} \sum_{K=1}^{K} \sum_{Q=1}^{QL_K} P_{rob}K * P_{rob}R_j * S_{KQ}^{rate} * \overline{X}_{KQ} \qquad (13)$$

Where, $\lambda_{stram}^{broad}$ denotes the arrival rate to the broadcast channels, $AS^{broad}$ denotes the average streaming rate of the dedicated channels and $\overline{X}_{KQ}$ denotes the completion of the $X_{KQ}$. Furthermore, as b is the available bandwidth, the number of streams that can be concurrently supported by the main server is calculated as follows: $N^{brod} = \frac{b - b^{brod}}{AS^{broad}}$, and in Equations 6 and 7, the overall blocking probability can be found accordingly [6][38].

## 5. CACHING MECHANISM OF THE PROXY SERVER CACHE SCHEME

In this paper, the proposed caching technique aims at minimizing the time delay that is caused when using the broadcasting technique. This type of delay happens once the client loses the video broadcast of the first segment. In order to view the video, the client must wait for the following video broadcast based on the first segment within a different or the same channel. In order to minimize the time delay for this video, a novel caching mechanism is produced where it is called in this paper as the Proxy Server Cache mechanism. Figure 6 shows a representation of a single VOD system within a service area where two clients (PDA 2 and PDA5) cache the first segment from the LPS once it is lost from the LF. When these clients lose the first segment being broadcasted by LF1, the clients are accordinglycached through to the first segment from LPS1. The remaining segments must be retrieved from the LF after switching into any LF logical broadcasting channels.

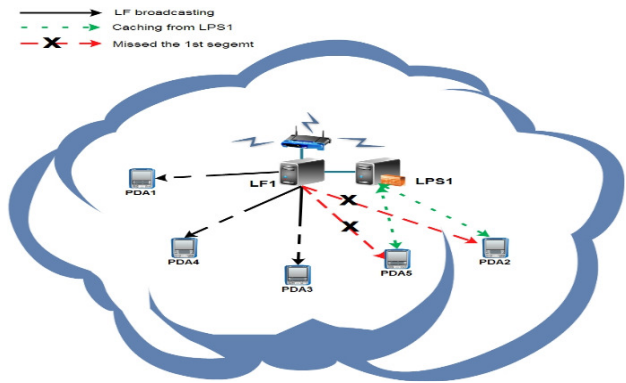

Figure 6. A representation of one VOD system service area

The novel mechanism is proposed based on the use of the LPS, which provides the first segment to the late clients who lose broadcasting the first segment once they arrive. The direct provision of the first segment saves the time of the clients. This can be conducted in order to avoid the time delay when the clients are waiting for the following video broadcast pertaining to the first segment. In this mechanism, every LF should pass a copy of the first segment of the videos through to the neighboring LPS based on the coverage of its area. Figure 7 illustrates the way on how the LF passes the first segment of the broadcasted videos through to the neighboring LPS.

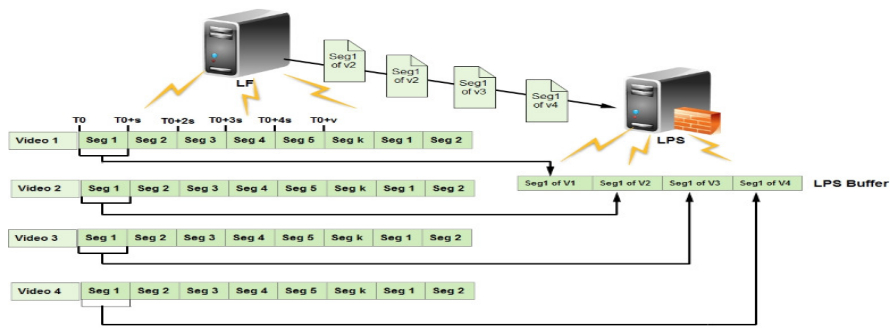

Figure 7. The Process of sharing the first segment videos of the LF with LPS

International Journal of Computer Networks & Communications (IJCNC) Vol.10, No.6, November 2018

### 5.1. MANAGING THE LOAD OF THE LPSS

In the proposed scheme, a high number of clients may request the video concurrently. The LPS in a VOD system has the ability to accommodate these requests. LPSs have to serve late clients by sending the first segments anytime. A mechanism should be presented to maintain the resources of these LPSs. In the proxy server caching scheme, every LF retains information regarding the LPSs, which are located within the coverage of its area. Table 1 represents the information that is used to manage the requests of the LPSs by assigning the requests to the LPSs in order to balance the requests among these LPSs for the sake of maintaining the resources of the LPSs. This information includes the name of the LPSs, the address of each LPS, the number of clients who are being served by the LPS and the IDs of the clients.

Table 1. The LPS Information of the LF Table

| Proxy Name | Proxy Info | No. of requests | Clients IDs |
|---|---|---|---|
| **LPS1** | IP address, port | 5 | C1, C3, C5, C7, and C9 |
| **LPS2** | IP address, port | 4 | C2, C4, C6 and C8 |

By using this mechanism, the VOD system ensures a fair distribution of requests or a balanced load between the LPSs where none of them has more load over the others. When a late client requests a video, the LF checks the table to identify which LPS has the least number of requests. After that, the LF retrieves information of the appropriate LPS and transmits it to that particular client. The client caches the first segment from the LPS that has the least number of clients. If all the LPSs have an equal number of requests, the LF sends the request to the LPS that has a smaller ID address number. By managing the load over the LPSs of the VOD system can work effectively when the number ofclients increases. This is due to the limitation of wireless transmission media.

## 6. SCENARIOS OF THE PLAYBACK VIDEO IN THE PROXY SERVER CACHE MECHANISM FOR THE VOD SYSTEM

In the proposed scheme, the client requests for a video after accessing one of the LF's areas. The LF checks the arrival time of the current client in order to differentiate between the cases on whether or not this client loses the first segment of the broadcasted video or not. The starting time of the broadcasted video is set to $T_0$. If the client arrives at $T_0$, it can join that channel and view the video. If the client arrives between the times $T_0$ and $T_0+s$ of the broadcasting time, the client will accordingly lose the first segment. Based on this case, the LF traces its table to determine which LPS is appropriate for streaming the video of the first segment through to the late client and transmits the information of the LPS to its client. After that, the client sends a request for the first segment to the LPS. After that, the client joins one of the LF's channels to obtain the rest of the segments. According to the schedule of the SB protocol, the mobile client arrives either at time $T_0$ or at a time between $T_0$ and $T_0+s$ of the SB scheduler. Therefore, there are two scenarios to playback the video in the proxy server caching mechanism. Sections 6.1 and 6.2 elaborate these two scenarios.

### 6.1 THE FIRST SCENARIO OF THE PLAYBACK VIDEO (V3) BY CLIENT J

In this scenario, Client J arrives at $T_0$ (the beginning of video (V3) broadcasting), so the procedure of V3 playback by Client J is performed according to the following steps:

**Procedure One of the Playback Video**

    **Begin**
  Client J accesses the LF area and requests V3.
    LF checks the arrival time to find Channel r that broadcasts the first segment of V3.
     *IF* the arrival time of Client J is listed in the LF
       Then, the LF informs the client about Channel r.
       Client J joins the Channel r.
    *Else* The LF will keep checks the arrival time
      *IF* the initial buffer and per-fetch buffer is available in the Client J buffer
- The Client J receives the first segment into the initial buffer.
- The Client J receives the rest of the segments into the per-fetch buffer.
- And then, Client J plays the first segment from the initial buffer, then switches to the per-fetch buffer and plays the remaining segments.
- Client J releases Channel r.

      *Else* The Client J cant receives segment into the buffer.
    **End.**

The first scenario when a Client J arrives at time $T_0$ of broadcasting the first segment of Video V3 on Channel r. Client J requests the first segment of V3 from the LF and the LF informs Client J that Channel r will broadcast the first segment accordingly. Hence, Client J has to join this channel. After that, Client J joins Channel r and starts streaming Video V3.

### 6.2. SECOND SCENARIO OF PLAYBACK VIDEO (V3) BY CLIENT J

In this scenario, Client J arrives at a time between time $T_0$ and $T_0+s$ (the case of missing the broadcast of Video V3 of the first segment).

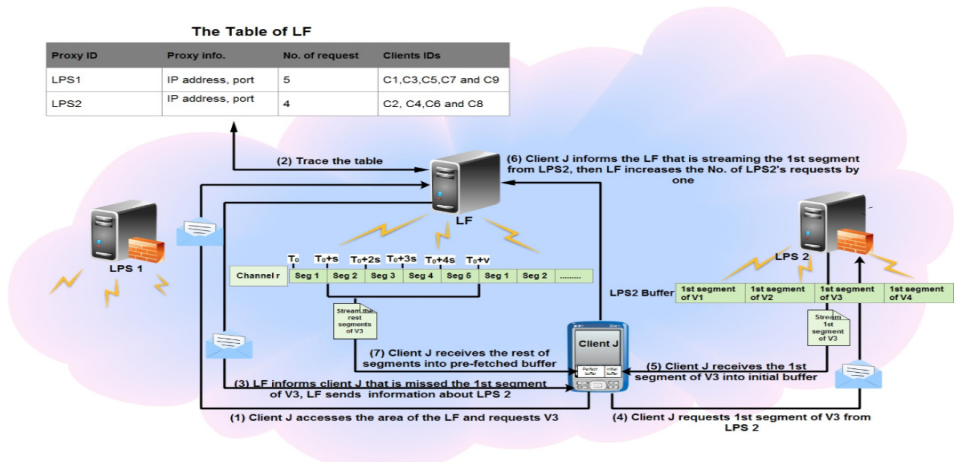

Figure 8. The second scenario of Video V3 playback by client J

The procedure of viewing Video V3 by Client J is performed according to the following steps:

**Procedure Two of the Playback Video**

   **Begin**
                Client J accesses the LF area and requests V3.
                The LF checks the arrival time in order to find the channel that will broadcast the first segment of Video V3.
    **IF** the arrival time of Client J is NOT listed in the LF.
- The LF informs Client J it loses the first segment on Channel r.
- The LF traces its table for a suitable LPS.
- The LF sends information about the LPS (*i*) that has the least number of requests.

        **IF** the Client J got the information about LPS from LF
- Client J requests the first segment of Video V3 from the LPS.

        **IF** the initial buffer and per-fetch buffer is available in the Client J buffer
- Client J receives the first segment of Video V3 into the initial buffer.
- Client J informs the LF that is streaming the first segment of V3 from LPS i in order to update the value of the number of LPSs' requests within its table.

        *Else* The Client J cant receives segment into the buffer.
- The LF increases the number of requests of LPS (*i*) by one.
- Client J joins Channel r.
- Client J receives the rest of the segments from channel r into the per-fetch buffer.
- Client J plays the first segment from the initial buffer then switches to the per-fetch buffer and plays the remaining segments.

        **Else** Client J keeps waiting the information about LPS
    **Else** Client J releases Channel r.
   **End.**

Figure 8 shows the second scenario of Video V3 playbacks when Client J arrives between the times ($T_0$ and $T_0+s$) where this client caches the first segment from the LPS. Client J requests Video V3 from the LF. The LF traces its table since Client J loses broadcasting the first segment from its channels. The table shows that LPS2 has a less number of requests, so the LF informs Client J that it loses broadcasting the first segment of V3 and it should request it from LPS2. After that, Client J sends back a message to inform the LF that it is currently obtaining the first segment of Video V3 from the LPS2. The LF increases the number of LPS2's requests by one. Finally, Client J joins Channel$_r$ in order to obtain the remainingsegments.

### 6.3 THE FULL PROCEDURE OF VIDEO V3 PLAYBACK BY CLIENT J

This section shows the complete procedure of Video V3 playbacks by Client J with two cases of the arrival time (either at T0 or between T0 and T0+s) to the VOD system as follows:

**Procedure 3 of the Playback Video**
   **Begin**
        Client J accesses the area of the LF and requests Video V3.
        The LF checks the arrival time of Client J and searches for a channel that will broadcast the first segment of VideoV3.
   **IF** the arrival time is between T0 and T0+s, then:
- The LF informs Client J that it has missed the first segment on Channel r.
- The LF traces its table to find the suitable LPS.
- The LF sends the information about the LPS (*i*) to Client J.
- Client J requests the first segment of VideoV3 from LPS (*i*).

       - LPS(*i*) pass the first segment to Client J.
       - Client J receives the first segment into the initial buffer.
      - Client J informs the LF that it is streaming the first segment from LPS i in order
      to update the value number of requests of LPS (*i*) in the LF table.
        - The LF increases the number of requests of LPS(*i*) by one.
      - Client J waits to join Channel rin order to obtain the rest of the segments.
    **Else IF** the arrival time is T0:
     - Client J joins Channel r.
     - Client J receives the first segment into the initial buffer and the rest segments
     into the pre-fetch buffer.
     -Client J receives the rest of the segments into the pre-fetch buffer.
     -Client J plays the first segment from the initial buffer, then switches to the per-
     buffer and plays the rest of the segments.
    **Else** Client J releases Channel r.
  **End.**

The video playback procedure is performed in the proxy server caching mechanism. This procedure states the full scenario of the arrival time of any clients. In the proposed scheme for the VOD system, the main server starts broadcasting the video to the LFs. The LF classifies the video into a number of segments and then broadcasts it over a number of logical broadcasting channels so that any new client can request the video and view it. The client may arrive in either of two cases. The first case emerges once the client loses broadcasting the first segment and the second case emerges when the client does not. If the client arrives at time $T_0$ when Channel r is broadcasting, the client can be included in this channel and streams the entire segments. If the client arrives at the time after the channels have started broadcasting the first segments, the client has to request the missed segment from the LPS (i) that is assigned by the LF. After that, the client requests the first segment from LPS i. Consequently, the LF increases the requests number of LPS (*i*) by one. The rest of the video segments are obtained from Channel r of the LF.

## 7. SYSTEM PARAMETERS

This section elaborates the parameters as shown in Table 2, which are used to evaluate the proposed Proxy-Cache scheme for theVOD system. The 60-minute-video of the MPEG-1 encoding is divided into K segments of an equal size. The number of segments depends on the number of logical broadcasting channels that are based on the bandwidth of transmission media. According to Equation 8 that is represented for the SB protocol, the length of each segment will be 12 minutes when the video length is set to 60 minutes and the number of logical broadcasting channels is equal to 5 logical channels. The number of segments of the SB protocol must be equal to the number of logical broadcasting channels. The number of these channels per each video is identified according to the bandwidth of the IEEE802.11g. According to the parameters of Equation 9, the bandwidth (*K*) of the transmission media (IEEE802.11g) is $B = 45\ Mbps$, and which is used to transmit the videos among the system devices. The consumption rate video is $r = 1.5$ of the MPEG-1 encoding. The maximum number of videos of the Proxy-Cache scheme for the VOD system that can be handled is N<=5 videos. By applying the parameters of Equation 2, the parameters are comprised as follows: $1.5 * K * 5 < 45$. The value of *K* should be less than 7 logical channels for each video in order to achieve the requirement of this formula. A better number of logical channels of each video can be *K*=5.Based on the SB protocol's scheduler, the number of segments related to every video should be equal to the number of logical channels, which are 5 segments allocated for each video. Every broadcasting channel broadcasts the video segments based on a transmission rate are less or equal to the consumption rate. This is in order to avoid the client to buffer a big portion of a video in its buffer during the streaming process. Table 2 illustrates the evaluation parameters of the Proxy-Cache scheme for the VOD system where the

International Journal of Computer Networks & Communications (IJCNC) Vol.10, No.6, November 2018

Wi-Fi IEEE802.11g (54 Mbps) technique is used to provide the video service to the clients. Five segments of the video are broadcasted over five logical broadcasting channels and at different video lengths of 30, 60 and 90 minutes in order to evaluate the delay of the service under these lengths. The consumption rate of the MPEG-1 encoding is 1.5 Mbps. Different numbers of clients arrive at different periods of time, where {1,2,3,4,5,6,7,8,9,10} clients arrive at each minute.

Table 2. Evaluation Parameters

| Parameters | Notation | Values |
|---|---|---|
| **System bandwidth** | B | 54 Mbps |
| **The number of logical broadcasting channels** | *K* | 5 logical Channels |
| **The number of video segments** | K | 5 Segments |
| **Video length** | L | {30, 60 and 90} minutes |
| **Video consumption rate** | V | 1.5 Mbps |
| **Client arrival rate** | C arrival | {1, 2, 3, 4, 5, 6, 7, 8, 9 and10} (Clients/minute) |
| **The number of videos** | N | 1-5 |

## 8. EXPERIMENTAL ENVIRONMENT

In this section, the equipment that is used for applying the PSCM caching scheme pertaining to the VOD system is presented in detail. The implementation of this scheme is divided into three classifications, which comprise verified tools, developing tools and hardware environments. More discussions regarding these classifications are highlighted in the following sections.

### 8.1. VERIFIED TOOLS

The VLC (Video LAN) software is an open source cross-platform that is used as a verified tool to test the behavior of the system and to ensure a proper stream for the video through the system devices. The server in the system handles the video to be streamed into multiple numbers of mobile clients by using the VLC framework. This framework has the capability to work with various types of video encoding, such asthe MPEG-2, H.264, WMV, DivX, MPEG-4and FLV. Additionally, it can efficiently work with several operating system platforms.

### 8.2. DEVELOPMENT TOOL

The tool that is being used to develop the Proxy-Cache scheme of the VOD system is the Microsoft Visual C# 2010. This development environment allows the programmer to have a flexible capability to deal with streaming the video though to the networks.

### 8.3. HARDWARE ENVIRONMENTS

This section represents the hardware devices that are used to evaluate the Proxy-Cache scheme of the VOD system. The implementation focuses on one LF service area where multiple mobile clients efficiently join the requested videos. Table 3 illustrates the environments that are located in the LF service area accompanied with the specifications of each of these hardware devices.

Table 3. Hardware Environments

| Types of Devices | No. of Devices | The Specification of the Device |
| --- | --- | --- |
| **Client Devices** | 5-10 | Tested on mobile devices (laptop), Inspiron 5559, Intel Core i7-6500 U Inside Processor 3Ghz, 4G Cache |
| **Local Forwarder (LF)** | 1 | Intel (R) Core(TM)2 Duo CPU T6600, 2.20Ghz |
| **Local Proxy Server(LPS)** | 2 | Stationary device, Intel Dual-Core AMD Opteron (tm), Processor 2214 2.20 GHz (2 Processors ) |
| **Operation Systems** | 13 | Windows 7, 64-bit OS |
| **Wireless Transmission** | 1 | Wi-Fi IEEE802.11g, 54Mbps, WIMAX, 72 Mbps |

## 9. EXPERIMENTAL RESULTS AND DISCUSSIONS

This section introduces the produced findings that are obtained by applying the proxy server cache scheme for the VOD system. Additionally, the service delay of ordering the video based on several effective cases is discussed. The time delay is considered to be the time average for which a client should begin from ordering the video until the recipient of its service. The time average measures how the VOD system functions as effectively through anon-demand video service.

### 9.1. THE EFFECTS OF THE NUMBER OF ARRIVAL REQUESTS WITH DELAY

The obtained findings of the time delay put an impact on the VOD system after its measurement relies on the number of clients' requests within a particular time as shown in Figures 9 (a) and 9 (b) . It is found to be proven from the obtained results that the average time delay of the five caching techniques comprises All-Cache, DSC-Cache, Random-Cache, Proxy-CacheandPoR-Cache. The average service delay without caching represents half of the duration of the first segment (V/K/2 = 60/5/2=6 minutes for 60 minutes of a video length). It is found to be proven from the results that the caching process helps to reduce the waiting time of the mobile clients substantially. In all types of caching, the system needs at least 2 clients to be situated within the area in order to cache the first segment and to reduce the time delay. Accordingly, the request arrival number starts from number two as the arrival request number. In all five caching techniques the client population is sparse ($P_{arrival=}$ 2), the delay is less than 90 seconds and when we test it in the ($P_{arrival=}$ 6), it is seen to be less than 40 seconds, which is 4 times better than without caching. These improvements are even more notable as the request rate increases. This is because when a client population becomes denser a client has a better chance to find a cache, thus reducing the service delay. In the new scheme, the time delay is considered effective regardless of the existence of a single client within a particular area. This scheme aims atobtaining a video service that has the same service for the entire clients. This implies that the client could obtain the first segment regardless there are previous clients within the indicated area or not. Proxy-Cache almost provides TVOD services since it offers a delay that is less than 5 seconds in most scenarios as shown below. When the arrival rate either increases or decreases, mobile clients can find the first segment smoothly. In the All-Cache, DCS-Cache, Random-Cache and PoR-Cache scheme, the system needs at least two clients to be situated in the area in order to cache the first segment and to reduce the time delay. Consequently, the request arrival number starts from number two as the arrival request number. In the Proxy-Cache scheme, the late client can cache the first segments regardless of the availability of other clients located in the area. Hence, the arrival request number starts from number one, which indicates to the late client who obtains the first segment regardless of the existence of previous clients within the area.

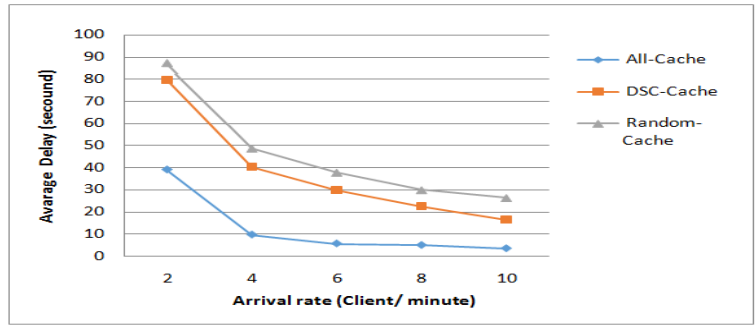

Figure 9 (a). The average time delay of the All-Cache, DSC-Cache and Random-Cache scheme based on the number of arrival requests (Client/minute)

As shown in the All-Cache scheme, the average delay of viewing a video is 5.61374 seconds, in the DCS 29.757682 seconds, and in the Random-Cache 37.672158 seconds when the number of arrival requests is ($P_{arrival}$= 6) per one second and PoR-Caching is 478.3601 milliseconds. Meanwhile, in the Proxy-Cache scheme, the average delay is 142.2766 milliseconds when the number of clients is set to 2. In the Proxy-Cache scheme, the service delay is 164.3577 milliseconds when 6 clients arrive at every minute. Moreover, the average delay in the Proxy-Cache scheme is 179.2505 milliseconds when 10 clients arrive every minute. The delay that occurs in the All-Cache, DCS and Random-Cache scheme is due to the reason that the late client has a less chance to search for a client who catches the same first segment of the video. In the Proxy-Cache scheme, the late clients have a higher chance to cache the same first segment of the video from the LPS. The results show that the average delay of the Proxy-Cache scheme is slightly increasing at a time the number of clients is increased. Nonetheless, the increase of the delay is still reasonable in comparison with other caching schemes.

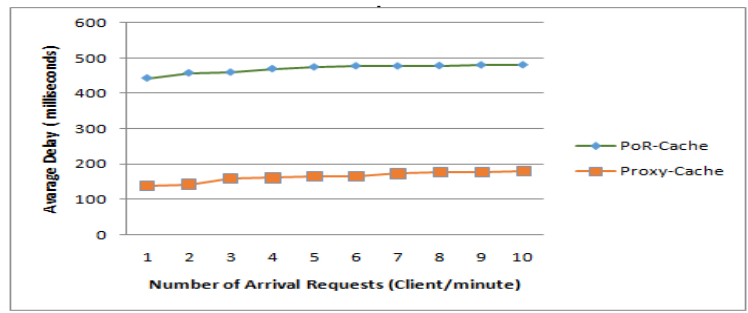

Figure 9 (b). The average time delay of the PoR-Cache and Proxy-Cache Scheme based on the number of arrival requests (Client/minute)

## 9.2. THE EFFECTIVE OF THE AVERAGE DELAY DEPENDING ON THE LENGTH OF THE VIDEO

The effect of the average time delay relies on several different video lengths where the VOD system is could manage different video lengths. Based on Equation 2, the length of every segment can be increased or reduced according to the total length of the video length and to the number of logical broadcasts. Figures 10 (a) and 10 (b) show the impacts of the video length on the service time delay through the All-Cache, DSC-Cache, Random-Cache, Proxy-Cache scheme and PoR-Cache, respectively. The average time delay reaches 5.362536 seconds when the video length reaches 60 minutes in the All-Cache scheme, 23.342363 seconds in the DSC-Cache, 35.303746 seconds in the Random-Cache and 476.2478 milliseconds in the PoR-Cache. The average service delay in the new scheme is 140 milliseconds. Further, the average service delay is 7 seconds when the video length reaches 90 minutes. The average service delay of the new scheme, in return, retains the same (i.e. 140 milliseconds) when the video length reaches 90 minutes. It can be inferred from the results that the average service delay of the All-Cache, DSC, Random-Cache and PoR-Cache schemes are increased by raising the length of the video. The increase in the delay is due to the fact that the size of the first segment, which caches from the neighbor, is increased. This, in turn, leads to disconnect the transmission when the cache is a mobile device. The transmission disconnection of a long segment causes an increase on the start-up overhead that is needed to find another new free client who has a cache of the same first segment video after the current cache has failed. However, when the failure rate increases, the start-up overhead will be increased as well. Thus, the longer length delay also increases. The results of the Proxy-Cache scheme show that the average service delay is considered to be the same when the length of the video increases. The reason of this is that the late client could directly cache the first segment from the stationary device where there is no start-up overhead to trace for another new cache after the current one moves away or terminates the system. Consequently, the Proxy-Cache scheme outperforms the failure in order to obtain the first segment so that there is no start-up overhead. Therefore, the delay of different video lengths maintains the same.

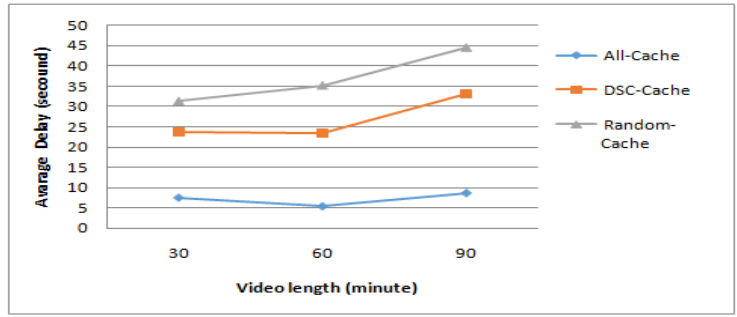

Figure 10 (a).The average delay of the All-Cache DSC and Random-Cache based on different video lengths

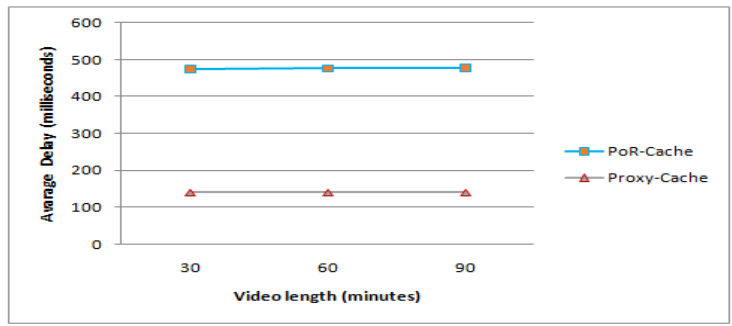

Figure 10 (b).The average delay of the PoR-Cache and Proxy-Cache scheme based on different video lengths (30, 60, and 90)

## 9.3. THE RESULTS OF THE FAIL PROBABILITY DEPENDING ON THE NUMBER OF ARRIVAL REQUESTS

The findings related to the failure probability that has the first segment video of the f caching schemes comprise the All-Cache, DSC-Cache, Random-Cache, PoR-Cache and the Proxy-Cache scheme based on the received request number. In the All-Cache scheme, a failure is identified when the new client fails to search for a free neighboring client who caches the first segment as shown in Figure11. Additionally, a failure is identified when the new clients fail to obtain the first segment of the video from the new cache. The reason behind this failure refers back to that the client who caches the first segment is found to be moving or is ended from the system.Figure11 illustrates comparative results of the failure probability for obtaining the first segment in the PoR-Cache and the Proxy-Cache schemes. Both schemes have close results to the newly proposed scheme. This probability of the PoR-Cache scheme is 0.0603173 and the All-Cache scheme as illustrated indicates 0.3903173 when the number of arrival requests is 2. At the same time, the failure probability for obtaining the first segment is 0 in the Proxy-Cache scheme, whereas the failure probability in the PoR-Cache is 0.01927233 and the All-Cache scheme is 0.03427786 when the number of arrival requests is 10. Meanwhile, the probability of failure in the Proxy-Cache scheme remains at 0. Since the probability of obtaining the first segment in the PoR-Cache and All-Cache schemes depends on the previous clients, the worst case of this failure arises whenever there is no any client who caches the first segment to the late clients. In the Proxy-Cache scheme, there is no any failure probability as the number of clients that is taken by the system is reasonable to the Wi-Fi bandwidth. Nonetheless, the probability of failure in the Proxy-Cache scheme is encountered when a high number of clients arrive at the same time and all of these clients attempt to obtain the first segment from the LPSs.

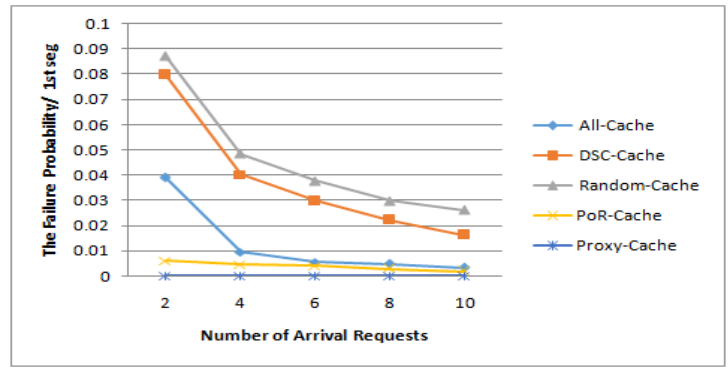

Figure 11. The findings of the failure probability when obtaining the first segment of the video and when the number of arrival request (2, 4, 6, 8 and 10)

## 10. CONCLUSION

The importance of this paper is to present a solution on how to minimize the time delay through the Video on Demand (VOD) system. Therefore, this paper proposes a new caching scheme in order to minimize the start-up delay in the VOD, which is called theProxy Server Caching Mechanism where the first segments of the entire broadcasting movies are kept in a stationary LPS server. If the client arrives after the first segment is broadcasted by the local forwarder, the clients will not obtain it. Hence, the client should wait for the following broadcasting channel, which can broadcast the first segment. In order to tackle this issue, aninstalling stationary LPS Server is suggested in this paper in order to ensurethat the late mobile client can obtain the first segment when it arrives with a less start-up time delay in comparison with the other schemes. This caching reduces the time delay by ensuring that the late client obtains the first segment once it arrives. In order to retain the resources of the LPSs within the system, a load balancing method is applied with the new scheme. The results show that the new proposed caching scheme is much better than other caching schemes where this is based on the impacts of the number of arrival requests with the time delay, the impact of the average time delay based on the length of the video and the findings of the failure probability based on the number of arrival requests.

International Journal of Computer Networks & Communications (IJCNC) Vol.10, No.6, November 2018

## AUTHOR

**Saleh Ali Alomari** obtained his MSc and Ph.D. in Computer Science from UniversitiSains Malaysia (USM), Pulau Penang, Malaysia in 2008 and 2013 respectively. He is a lecturer at the Faculty of Science and Information Technology, Jadara University, Irbid, Jordan. He is Assistance Professor at Jadara University, Irbid, Jordan 2013. He was the head of the Computer Network Department at Jadara University from 2014 until 2016. He is the candidate of the Multimedia Computing Research Group, School of Computer Science, USM. He is research assistant with Prof. Dr. Putra, Sumari. He is managing director of ICT Technology and Research and Development Division (R&D) in D&D Professional Consulting Company, Malaysia. He has published over 40 papers in international journals and refereed conferences at the same research area. He is a member and reviewer of several international journals and conferences (IEICE, ACM, KSII, JDCTA, IEEE, IACSIT, etc). His research interest is in the area of multimedia networking, video communications system design, multimedia communication specifically on Video on Demand system, P2P media streaming, MANETs, caching techniques and for advanced mobile broadcasting networks as well.

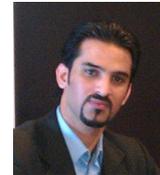